\documentclass[]{spie}  

 \def\farcs{\hbox{$.\!\!^{\prime\prime}$}}

\newcommand\arcsec{\mbox{$^{\prime\prime}$}}%
 
\usepackage{amsmath,amsfonts,amssymb}
\usepackage{graphicx}
\usepackage[colorlinks=true, allcolors=blue]{hyperref}
\usepackage{caption}
\usepackage{subcaption}

\title{On-sky performance and recent results from the Subaru coronagraphic extreme adaptive optics system\thanks{~~Based on data collected at Subaru Telescope, which is operated by the National Astronomical Observatory of Japan.}}

\author[a,b]{Thayne Currie}
\author[b,c,d,e]{Olivier Guyon}
\author[b]{Julien Lozi}
\author[b]{Ananya Sahoo}
\author[b,e]{Sebastien Vievard}
\author[b,f]{Vincent Deo}
\author[g]{Jeffrey Chilcote}
\author[h]{Tyler Groff}
\author[i]{Timothy D. Brandt}
\author[j]{Kellen Lawson}
\author[b,f,k]{Nour Skaf}
\author[l]{Frantz Martinache}
\author[m]{N. Jeremy Kasdin}
\affil[a]{NASA-Ames Research Center, Moffett Field, California, USA}
\affil[b]{Subaru Telescope, 650 N. Aohoku Pl., Hilo, Hawai'i, USA}
\affil[c]{Steward Observatory, University of Arizona, Tucson, AZ, USA}
\affil[d]{College of Optical Sciences, University of Arizona, Tucson, AZ, USA}
\affil[e]{Astrobiology Center of NINS, 2-21-1, Osawa, Mitaka, Tokyo, Japan}
\affil[f]{LESIA, Observatoire de Paris, Université PSL, CNRS, Sorbonne Universit\'e, Universit\'e de Paris, 5 place Jules Janssen, 92195 Meudon, France}
\affil[g]{University of Notre Dame, South Bend, IN, USA}
\affil[h]{NASA-Goddard Space Flight Center, Greenbelt, MD, USA}
\affil[i]{University of California-Santa Barbara, Santa Barbara, CA, USA}
\affil[j]{University of Oklahoma, Norman, Oklahoma, USA}
\affil[k]{Department of Physics and Astronomy, University College London, London, United Kingdom}
\affil[l]{Univ. Cote d'Azur, Nice, France}
\affil[m]{University of San Francisco, San Francisco, USA}

\authorinfo{Further author information: (Send correspondence to T. Currie)\\T. Currie: E-mail: currie@naoj.org}

\begin{document} 
\maketitle

\begin{abstract}
We describe the current on-sky performance of the Subaru Coronagraphic Extreme Adaptive Optics (SCExAO) instrument on the Subaru telescope on Maunakea, Hawaii.   SCExAO is continuing to advance its AO performance, delivering H band Strehl ratios in excess of 0.9 for bright stars.   We describe new advances with SCExAO's wavefront control that lead to a more stable corrected wavefront and diffraction-limited imaging in the optical, modifications to code that better handle read noise suppression within CHARIS, and tests of the spectrophotometric precision and accuracy within CHARIS.   We outline steps in the CHARIS Data Processing Pipeline that output publication-grade data products.  Finally, we note recent and upcoming science results, including the discovery of new directly-imaged systems and multiwavelength, deeper characterization of planet-forming disks, and upcoming technical advances that will improve SCExAO's sciencec capabilities.
\end{abstract}

\keywords{adaptive optics, extrasolar planets, infrared}

\section{Introduction}
\indent In the past twelve years, facility adaptive optics (AO) systems on 8-10m class telescopes and now successor \textit{extreme} AO platforms have provided the first images of (super-)jovian extrasolar planets orbiting nearby, young stars\cite{Marois2008,Marois2010,Lagrange2010,Currie2014,Currie2015,Macintosh2015,Keppler2018}.  Depending on the exact adopted definition for a planet vs. a brown dwarf, about 12--20 directly-imaged planets have been discovered.   Nearly all imaged exoplanets have masses exceeding 5 jovian masses ($M_{\rm J}$) and orbit beyond 10--30 au from their host star.   Typical planet-to-star contrasts in the near-infrared range from 10$^{-4}$ to 10$^{-6}$; on the sky, projected separations range from 0\farcs{}1 to 1\arcsec{}.  

Recent extreme AO direct imaging surveys from the \textit{Gemini Planet Imager} (GPI)\cite{Macintosh2014} on Gemini-South and \textit{Spectro-Polarimetric High-contrast Exoplanet REsearch instrument} (SPHERE)\cite{Beuzit2019} on the Very Large Telescope (VLT) show that superjovian planets at 10--100 au are rare, but the frequency of these planets may increase closer to 5 au, consistent with a turnover in the jovian planet population suggested from radial-velocity surveys\cite{Nielsen2019,Fernandes2019}.  Photometric and spectroscopic follow-up observations of directly-imaged planets reveal key features in young jovian exoplanet/substellar atmospheres, including the presence of copious dust and thicker clouds from near-IR colors, non-equilibrium carbon chemistry from near-IR spectra and mid-IR photometry, low surface gravities from near-IR spectra, and molecular abundances from medium-resolution spectra \cite{Currie2011,Galicher2011,Rajan2017,Wilcomb2020,Molliere2020}.  

Longer-term, ground based extreme AO systems on the next generation of 30-m class telescopes 
are designed with the goal of directly imaging planets in reflected light around mature stars, including Earth-like habitable zone worlds\cite{Guyon2018}.
However, current ground-based capabilities only probe the extremes of planet formation: moderate to wide-separation superjovian planets.   The best near-IR post-processed contrasts achieved thus far -- $\sim$ 10$^{-6}$ at 0\farcs{}2 and 5$\times$10$^{-7}$ at 0\farcs{}5\cite{Vigan2015} -- are still 10--1000 times too bright to detect reflected-light jovian-sized planets around the nearest mature stars 
or self-luminous Saturn to Jupiter-mass planets around more distant young Suns on Jupiter-to-Neptune-like orbits.   Detecting and then characterizing solar system-like planets from the ground requires significant advances over capabilities demonstrated with the first generation of extreme AO systems.  

  In this paper, we describe the status, recent hardware/software advances, and recent science results for the Subaru Coronagraphic Extreme Adaptive Optics project (SCExAO) at the Subaru Telescope on Maunakea \cite{Jovanovic2015} coupled with its main near-IR planet-imaging science instrument, the CHARIS integral field spectrograph \cite{Groff2016}.  Section \S{2} motivates and describes wavefront control within SCExAO, noting recent changes that have improved its performance stability and correction for low-order modes.   Section \S{3} summarizes CHARIS, notes current its read noise level and changes to the data reduction pipeline to work around these problems, notes spectrophotometric calibration and astrometric calibration, and performance.   
  
  Section \S{4} provides a walkthrough of CHARIS data reduction steps currently adopted within the CHARIS data processing pipeline (DPP) \cite{Currie2018a}.   
  Section \S{5} gives a summary of recent science results, focusing on newly discovered companions to nearby stars and detailed characterization of planet-forming disks.   Section \S{6} outlines updated future directions for SCExAO and for CHARIS, describing upgrades that will improve the instrument combination's performance, enhance its planet detection capabilities, and broaden its science capabilities.  
  
      \begin{figure}[h]
   \begin{center}
   \includegraphics[scale=0.65,clip]{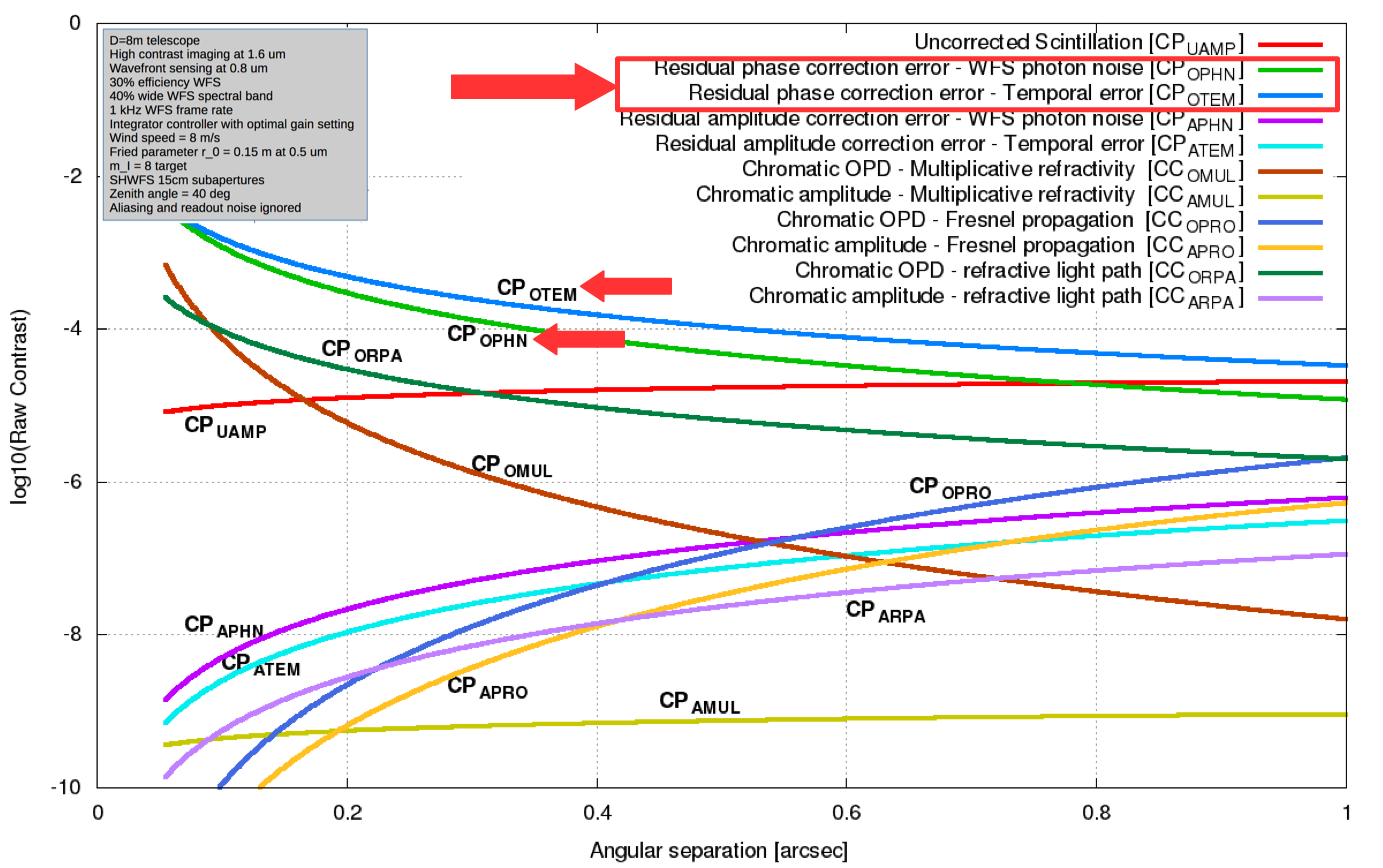}
   \vspace{-0.125in}
   \end{center}
   \caption
   { \label{fig:wfechallenge} 
   Wavefront error budget for a spatially-filtered Shack-Hartmann wavefront sensor running at 1 kHz and correcting turbulence for an $I$ = 8 star on an 8m telescope.   See \cite{Guyon2005,Guyon2018} for a more in-depth discussion.}
   \end{figure}
   
\section{SCExAO: Motivation, Architecture, and Recent Advances}
\subsection{The Wavefront Control Challenge}
Figure \ref{fig:wfechallenge} sketches an example wavefront error budget that illustrates the technical challenges imaging planets below the 10$^{-6}$ level from the ground with 8m class telescopes.   At small angles, contrasts are nominally most impacted by the temporal bandwidth error (``servo lag") and photon noise on the wavefront sensor (WFS). These two terms are also coupled: a faster-sampling WFS loop reduces temporal errors, as the WFS loop runs faster relative to the atmospheric coherence time, but suffers increased photon noise error.   Chromatic wavefront errors and scintillation further impede raw contrasts below 10$^{-5}$.   

Demonstrated contrast gains using advanced post-processing techniques range between a factor of 10 and 100\cite{Bailey2016}.   Thus, reaching post-processed contrasts approaching 10$^{-7}$ requires raw contrasts below 10$^{-5}$ and, in turn, significant reduction of dominant coupled error terms and then some suppression of chromatic wavefront errors.  SCExAO's wavefront control architecture is designed to significantly mitigate coupled temporal bandwidth and photon noise error terms, while providing a path forward to reducing chromatic errors.

\begin{figure}[ht]
   \begin{center}
  \centering
   \includegraphics[scale=0.33,clip]{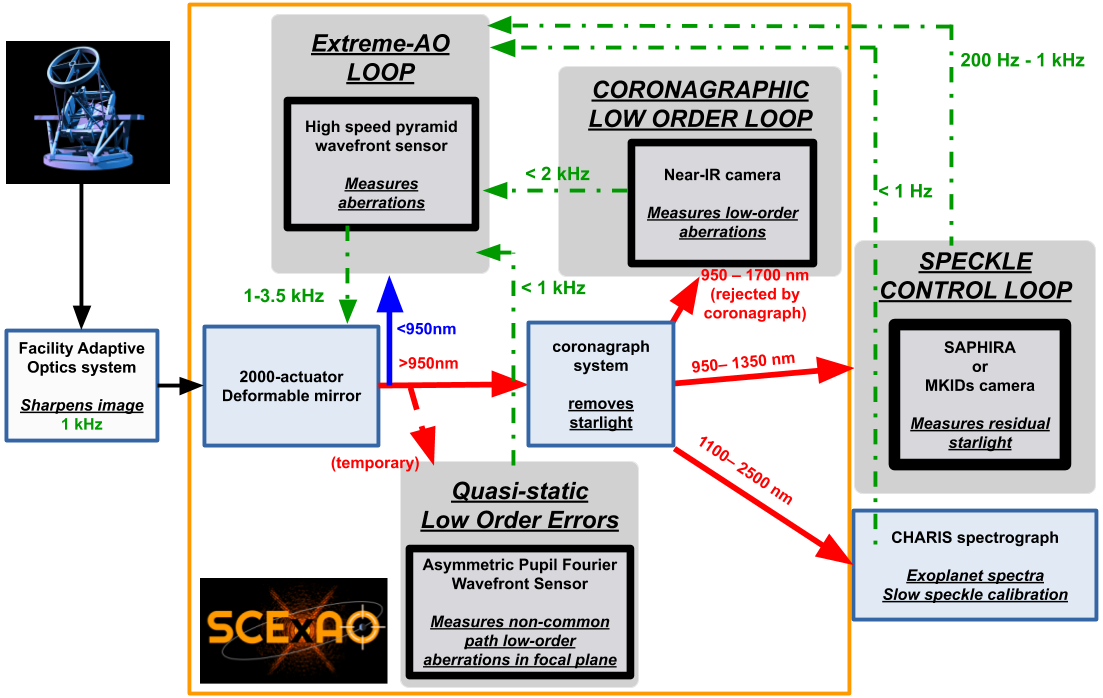}
   \end{center}
   \caption
   { \label{fig:wfcloops} 
       The current schematic of SCExAO.  Note that the coronagraphic low-order loop is not in normal operation; the MKIDs camera (MEC) is undergoing commissioning.
}
\end{figure}
\subsection{SCExAO Architecture} 


Figure \ref{fig:wfcloops} displays a schematic of high-contrast imaging with SCExAO.  
Incoming starlight is partially corrected for atmospheric turbulence using AO-188, Subaru's facility adaptive optics system, which typically achieves $\sim$ 30--40\% Strehl at 1.6 $\mu m$ under median to good conditions.   Then it is further sharpened by the SCExAO wavefront loop, reaching typical $H$ band Strehl ratios of 80--90\% under median to good conditions, with higher values for the brightest stars under the best conditions\cite{Currie2019b}.   The PSF core from sharpened starlight is then blocked and Airy rings suppressed by a suite of coronagraphs, where the standard Lyot coronagraph is typically used for observations requiring data over multiple bandpasses.   Finally, this light is fed into a suite of science instruments, where the CHARIS integral field spectrograph is the workhorse instrument operating over the $JHK$ bandpass.   

Common observing modes include sending $Y$ band light to a high speed broadband camera or MKIDS-based detector (MEC) and $JHK$ to CHARIS.    Alternatively, the broadband camera/MEC can take a broader bandpass and CHARIS focus on a single filter (e.g. $H$ or $K$).  For science cases focused on circumstellar material (e.g. protoplanetary disks) targeting optically bright stars, the bluer ($<$800~nm) optical wavelengths normally sent to the WFS (see below) are redirected to the VAMPIRES instrument\cite{Norris2020}.

\subsection{Wavefront Control with SCExAO: Recent Advances}
SCExAO's wavefront control (WFC) loop consists of a 2000-element MEMS deformable mirror from Boston MicroMachines driven by a modulated Pyramid wavefront sensor (PyWFS) using a double roof pyramid prism optic and an OCAM$^{2}$K EMCCD camera from First-Light Imaging that operates over a 600--900 nm bandpass (smaller if used in combination with VAMPIRES).   The SCExAO WFC loop can run at speeds up to 3.5 kHz.  However, for most science observations it operates at a slower 2 kHz for bright stars coupled with predictive wavefront control\cite{MalesGuyon2018} (see below) to decrease temporal bandwidth error and wavefront sensor noise, limited by an overall latency of just under 1 ms.   For very faint stars($I$ $>$ 9), we typically run the loop at 1 kHz.   The loop can correct for up to 1200 modes of dynamic aberrations.   

To operate SCExAO's wavefront control loop, we use the Command and Control for Adaptive Optics (CACAO) open-use software framework\cite{Guyon2018b}: see Paper 11448-145 for details.   CACAO uses a unified shared memory structure to read in images from multiple cameras and then issue DM commands and apply various offsets to the PyWFS.    SCExAO leverages the very high bandwidth of its MEMS DM, coupled to a precise calibration of the hardware latency, to acquire AO response matrices (RM) -- influence functions that map actuator offsets to changes in the focal plane --  at a significant speed.   
A complete sequence of Hadamard modes can be probed in just 1.25~s at the usual framerate of 2~kHz. Stabilized RMs with optimized signal-to noise ratio (SNR) are usually completed
within $\sim$10 minutes.

The bandwidth of the DM enables us to probe the WFS response faster than the typical turbulence fluctuations on sky, and SCExAO has had the capability to acquire or refine RMs on sky (e.g. "RM bootstrapping") for many several years.
This is ideal, as nonlinear effects from the PyWFSs\cite{Deo2019b} can be measured in-situ and compensated for.   Under the best seeing conditions (e.g. $\theta_{\rm V}$ $\sim$ 0\farcs{}15--0\farcs{}4), on-sky RMs result in extremely well-corrected point-spread functions, with estimated $H$-band Strehl ratios reaching 94\% for bright stars and extreme AO corrections even for stars as faint as R = 11.6 \cite{Currie2019a,Currie2019b}.  

However, with PyWFSs nonlinearities being quite volatile, such a strategy is only paying for much better-than-median seeing and temporally stable conditions. Otherwise, the measured RM is mostly filled with randomized nonlinear components from the PyWFS, and the control capability is reduced to $\sim$500 modes, with reduced stability.   The reliability of on-sky RMs to compute low-order modes is further compromised by the tendency of AO-188 to habitually split PSF cores or first Airy rings and the ``low-wind effect"\cite{Bos2020}.

\begin{figure}[ht]
   \begin{center}
  \centering
   \includegraphics[width=0.925\textwidth,clip]{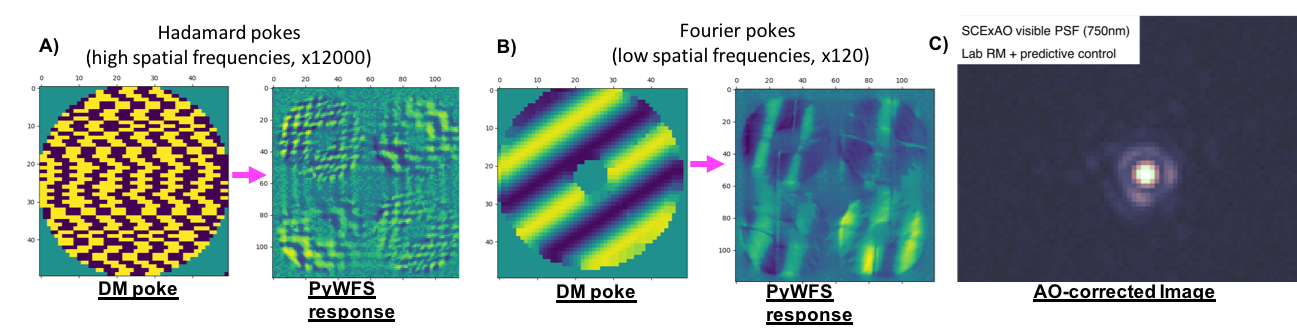}
   \end{center}
   \vspace{-0.18in}
   \caption
   { \label{fig:labrm} 
   Laboratory-based response matrix calculations for SCExAO blend a) Hadamard pokes sampling high spatial frequencies and b) Zernike/Fourier pokes sampling low spatial frequencies.   This approach, when combined with predictive control and in lieu of acquiring an RM on sky, yields far superior control of low-order modes and diffraction-limited imaging in the red optical (panel c).   Note that the rotation angle for the DM and PyWFS are offset by 45$^{o}$, explaining the tilt of the poke vs. PyWFS response patterns.
}
\end{figure}

Thus, we now prefer using command laws calibrated in the lab, coupled to frequency-dependent modal gain optimization.  As shown in panels a) and b) of Figure \ref{fig:labrm}, our lab RM calculation blends a high spatial frequency Hadamard poke sequence sensitive to high-order aberrations (12,000 pokes) with low spatial frequency poke patterns consisting of an explicit (non-orthogonal) mix of the first few Zernike modes and Fourier mode pokes up to three cycles per aperture (120).  While Hadamard pokes sequences provide a high SNR estimate of the WFS response at high spatial frequencies, they perform much more poorly at sampling low spatial frequencies.    Explicit Fourier/Zernike modes  compensate for these shortcomings.   

This strategy has considerably improved AO stability, and enables a significant fraction of ``salvageable time'' during nights with poor seeing and high wind speed. Noiseless, daytime RMs -- which lack non-linearities -- also provide a more stable baseline for on-sky training of predictive control\cite{GuyonMales2017}.
The raw broadband contrast for moderately bright stars ($I$ $\sim$ 5-7) under ``good" conditions (e.g. $\theta_{\rm V}$ $\sim$ 0\farcs{}5) using the lab-calibrated RM approaches and sometimes exceeds the best performance SCExAO has achieved on 1st and 2nd magnitude stars using an on-sky RM taken under exceptional conditions ($\theta_{\rm V}$ $\sim$ 0\farcs{}18--0\farcs{}45).  

While formally an on-sky RM could provide better sensing and control of the highest order modes, this advantage appears to be mitigated by the lab RM's superior low-order control, which allows the loop to be run at a higher gain, combined with improved efficacy of our coronagraphs, whose performance is particularly sensitive to low-order aberrations like tip-tilt (see \cite{Currie2018b}).   The visible PSF demonstrates that the lowest order modes are in fact exceptionally well controlled (Figure \ref{fig:labrm}, right panel).   In good conditions, the visible PSF is diffraction limited with a well defined PSF core and first Airy ring.

Finally, predictive control using empirical orthogonal functions\cite{Males2018} is now part of SCExAO's normal operation. Predictive control allows us to reduce the servo lag error in the wavefront sensing error budget, improving contrast at small angles, while running at a slightly slower 2 kHz loop speed.   
As shown in Figure \ref{fig:labrm} (panel c), predictive control improves PSF stability and contributes to diffraction limited imaging over a wide wavelength range.

\section{The CHARIS Integral Field Spectrograph}
\subsection{Overview}
CHARIS is a lenslet-based cryogenic integral field spectrograph capable of operating from 1.1 to 2.4 $\mu m$\cite{Groff2016}.   After receiving light that is well corrected from SCExAO and partially suppressed by a coronagraph, a sparse image is formed on the lenslet array.   After a pinhole array mitigates lenslet diffraction, light from the lenslet array is dispersed from one of two prisms onto a 2048x2048-pixel Hawaii 2RG detector into 135x135 30 pixel-long microspectra.   With the low-resolution prism in position, CHARIS spectra have a resolution of $\mathcal{R}$ $\sim$ 20 and cover 22 channels with central wavelengths of range 1.15--2.37 $\mu m$.   In its high-resolution mode ($\mathcal{R}$ $\sim$ 70), CHARIS spectra cover the J, H, or K passbands.   The CHARIS Data Reduction Pipeline (DRP) converts raw CHARIS detector data into data cubes ($x$ by $y$ by $\lambda$). 

\subsection{Characterization and Calibration of CHARIS Data}
\begin{figure}[ht]
   \begin{center}
  \centering
   \includegraphics[scale=0.425,clip]{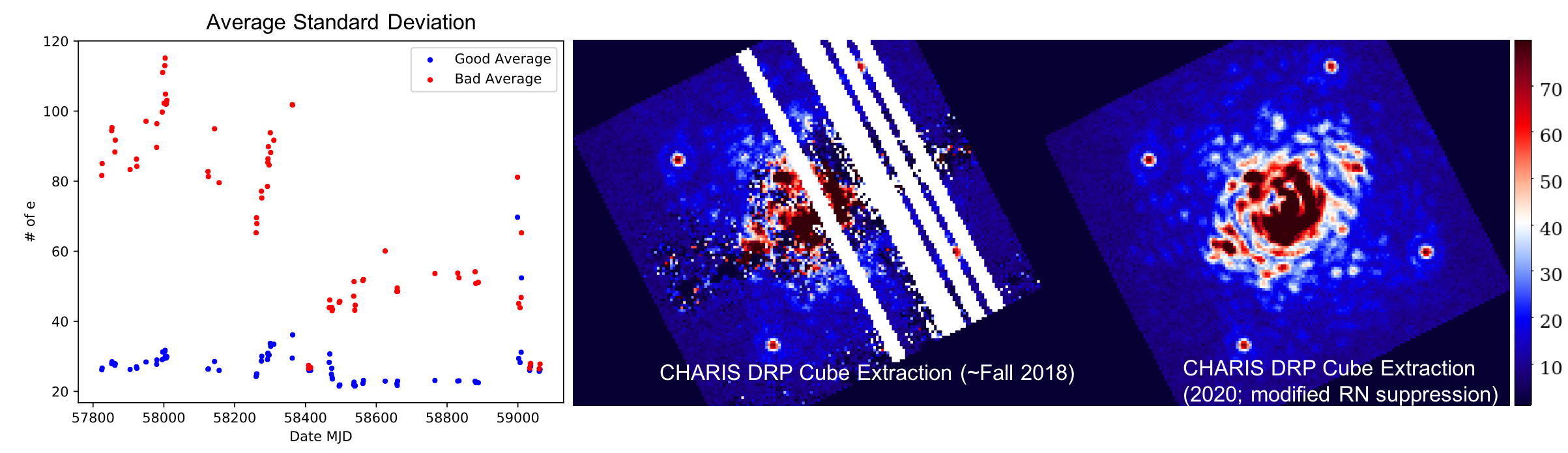}
   \end{center}
   \vspace{-0.15in}
   \caption
   { \label{fig:readnoise} 
       (left) Average standard deviation of pixels in low read noise channels (``good") and high read noise channels (``bad") as a function of time.   (middle) CHARIS data cube slice in $K$ band extracted from Fall 2018 data with elevated read noise, showing the failure of previous read noise suppression settings.   (right) The same cube with read noise suppression aggressiveness reduced.}
\end{figure}
\subsubsection{Read Noise}
Common among Hawaii 2RG detectors, CHARIS's detector suffers from $1/f$ read noise correlated among readout channels.   CHARIS's correlated read noise on the Nasmyth platform at Subaru is significantly higher than it was during pre-shipment laboratory testing at Princeton, enough to compromise the instrument's science capabilities\cite{Brandt2017}.    To suppress this read noise, the $\chi^{2}$ cube extraction method constructs and then subtracts from the detector data a two-dimensional model of the entire detector readout.   These residuals are then used to model and thus remove the correlated component of the read noise to levels approaching that of laboratory values.    

As shown in Figure \ref{fig:readnoise}, the typical CHARIS read noise level has fluctuated wildly for a subset of channels.   No one single cause for this change has been pinpointed, although the movement of CHARIS and other instruments on the Nasmyth platform may be connected to some fluctuations.    Our October 2018 run immediately followed an observatory power outage and warm-up of CHARIS.   The DRP's cube extraction program's read noise suppression failed, leaving swaths of the CHARIS image plane with exceptionally noisy (unusable) pixels along one axis and NaN stripes along a perpendicular axis (middle panel) for over 90\% of our exposures.   While elevated noise levels dropped to slightly lower levels during the next run, they reemerged in Summer 2020.   

To compensate, we relaxed the outlier rejection thresholds in \texttt{fitramps.pyx} to 30-$\sigma$ (line 603) and 20-$\sigma$ (line 609).    As shown in Figure \ref{fig:readnoise} (right panel), this modification recovered CHARIS data.   Comparing results using these new threshholds and previous ones for exposures with lower read noise levels showed negligible differences in data quality.  In addition to modifying the source code to adjust relax read noise suppression, we pointed the DRP to a different URL for populating fits header metadata responsible for determining the parallactic angle.  The previous one is down for maintainence indefinitely.   Specifically, in \texttt{calc$\_$metadata.py}, we added the following two lines in succession at the end of ``imports": \\
1) from astropy.utils import iers \\
2) iers.Conf.iers$\_$auto$\_$url.set('ftp://cddis.gsfc.nasa.gov/pub/products/iers/finals2000A.all')

CHARIS users are \underline{strongly encouraged} to verify their installation's settings for read noise suppression and the URL for fits header metadata and then modify them as described above, if needed.

\subsubsection{Spectrophotometric Precision and Accuracy}
\begin{figure}[ht]
   \begin{center}
   \vspace{-0.2in}
  \centering
   \includegraphics[width=0.32\textwidth,trim=14mm 0mm 12mm 0mm,clip]{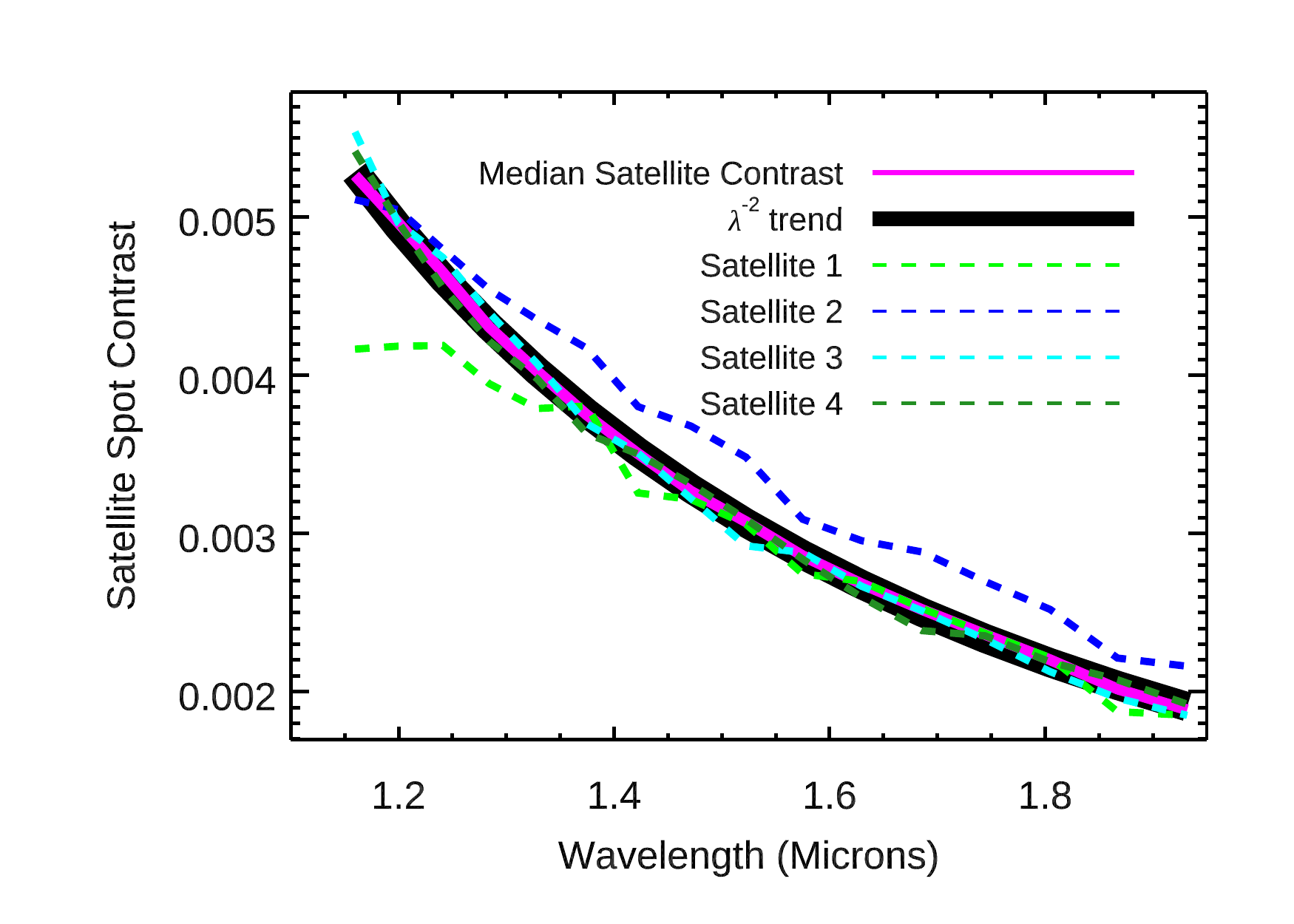}
   \includegraphics[width=0.32\textwidth,trim=14mm 0mm 12mm 0mm,clip]{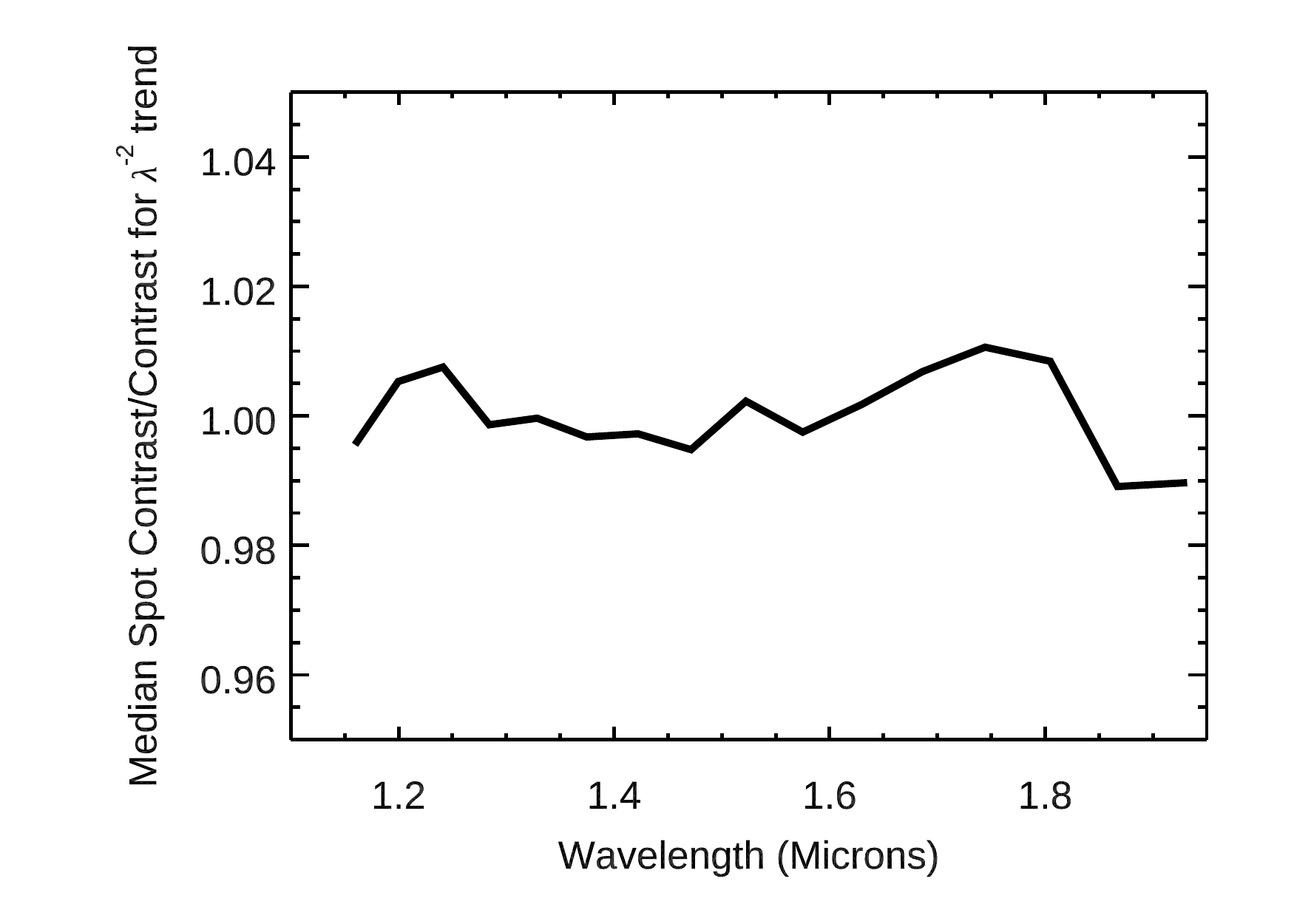}
   \includegraphics[width=0.33\textwidth,clip,trim=1mm 0mm 1mm 4mm,]{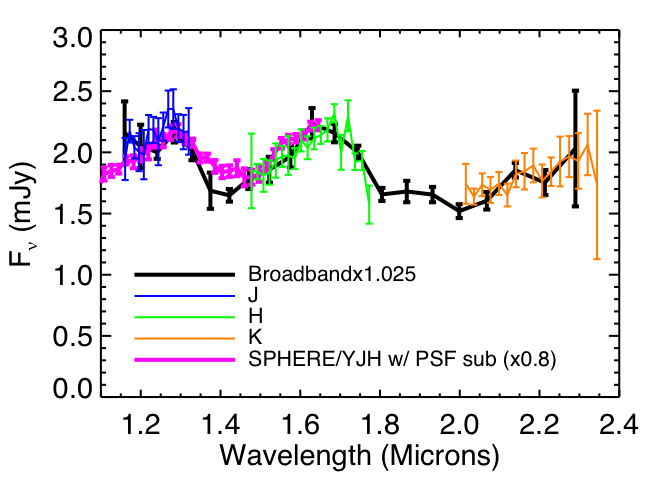}
   \end{center}
   \vspace{-0.2in}
   \caption
   { \label{fig:hd1160} 
       (left) Satellite spot contrast measured with CHARIS using the SCExAO internal source (1.1--1.8 $\mu m$) compared to the predicted $\lambda^{-2}$ trend, (middle) ratio of the median spot contrast vs. the predicted trend, and (right) Comparison between HD 1160 B spectra extracted from short CHARIS sequences in broadband, J, H, and K without PSF subtraction vs. a scaled VLT/SPHERE spectrum from Ref. \citenum{Maire2016} with PSF subtraction.   Our $JH$ photometry in high resolution and broadband mode ($J$=14.65 $\pm$ 0.07 and 14.71 $\pm$ 0.06, $H$ = 14.26 $\pm$ 0.07 and 14.28 $\pm$ 0.05) agrees with previous SCExAO/HiCIAO values within errors.
       }
\end{figure}

To perform spectrophotometric and astrometric calibration, CHARIS uses artificial satellite speckles. 
These speckles are generated by applying sine-waves to the DM and are unocculted copies of the central PSF.  Their intensity scales as $\lambda^{-2}$.   Using tests with SCExAO's internal source over a narrow bandpass centered on 1.55 $\mu m$ in Summer 2017, T. Currie measured the contrast of the speckles to be 6.412 $\pm$ 0.05 mags (2.72$\times$10$^{-3}$ $\pm$ 1.3$\times$10$^{-4}$) for a 25nm modulation.  The predicted contrast for a 50nm modulation should be exactly a factor of 4 higher (4.91 mags), which is completely consistent with our measurements: 4.92 $\pm$ 0.05 mag (1.08$\times$10$^{-2}$).   

Separate tests performed by J. Lozi in Fall 2018 used the superK laser in combination with CHARIS with the Lyot focal plane mask in place, simulating a CHARIS observation (Figure \ref{fig:hd1160}, top-left panel).    At 1.55 $\mu m$, the estimated contrast is $\sim$ 2.94$\times$10$^{-3}$, or $\sim$ 8\% higher.   The median-average of spot contrasts nearly perfectly matches the predicted $\lambda^{-2}$ trend, with a deviation of $\sim$ 1\% or less for each channel.    The contrasts for two of the four satellite spots show extremely small deviations from the median ($\sigma(spot/median)$ $\sim$ 2.2--2.4\%).  While the two other spots show slightly larger deviations when the full wavelength range is considered (5--6.4\%), over a smaller range ($\lambda$ $>$ 1.25 $\mu m$) their residuals are nearly the same as for the other two spots (2.6--3.2\%).   

Both tests were performed using the internal source where we differenced exposures with and without satellite spots to remove background speckles.   On sky under normal operation, the satellite speckles are always on.   Background speckles therefore limit the SNR of the satellite speckles and thus the precision of our spectrophotometric calibration.   The speckles lie at $\sim$ 15.9 $\lambda$/D in each channel, at the edge of the region well corrected by the DM.   In median to good seeing conditions when observing a bright star, the satellite speckles with a 25nm modulation typically have SNR $\sim$ 25--40 for most channels but lower for fainter stars, in channels covering telluric features, or at the reddest $K$-band channel where the thermal background is highest.

The righthand panel of Figure \ref{fig:hd1160} compares the CHARIS spectrum for for the HD 1160 B brown dwarf \cite{Nielsen2012} in broadband mode and in higher-resolution $J$, $H$, and $K$ bands to the published spectrum from \cite{Maire2016}.   No PSF subtraction techniques were applied to the CHARIS data, the total sequences spanned just a few minutes each, and the satellite spots were used only for the first few CHARIS data cubes for each band.    We adopted the 2017 spot absolute calibration.   The CHARIS broadband and higher-resolution spectra agree extremely well with one another modulo a slight (2.5\%) absolute calibration offset, which could be explained by slight changes in the AO performance or transmission between the satellite-on exposures and those lacking the satellite spots.  While the shape of our spectra agree with that for SPHERE/IFS outside of telluric-dominated passbands, our spectra are $\sim$ 20\% brighter.   However, our photometry integrated over the $JH$ passbands agrees far better with published values using unsaturated SCExAO/HiCIAO images\cite{Garcia2017} and also with subsequent SPHERE long-slit spectra\cite{Mesa2020}.  Photometry adopting the 2018 superK-derived absolute calibration agrees marginally less with SCExAO/HiCIAO photometry but is still consistent within errors.  

In summary, CHARIS can offer extremely good spectrophotometric precision and accuracy.   The median value of the satellite spots per channel follows the predicted $\lambda^{-2}$ trend to within 1\%;  the contrast scaling vs. modulation amplitude is also consistent with expectations.    However, we note a slight offset in absolute spectrophotometric calibration on order of $\sim$ 8\% between separate tests.   This difference is plausibly due to slight changes in the placement of SCExAO's focal-plane masks between 2017 and 2018, a parameter that will change on sky if the star is reacquired or observing suffers stretches of significant tip-tilt (e.g. before and after a transit near zenith).   Thus, achieving absolute calibration necessary to study variability at the $<$5\% level is not yet feasible with SCExAO/CHARIS.    Finally, the intrinsic SNRs of the satellite speckles used to flux-calibrate data cubes limit the precision of extracted spectra to $\sim$ 3--5\% precision per channel.
\begin{figure}[h]
\vspace{-0.1in}
   \begin{center}
  \centering
   \includegraphics[scale=0.65,clip]{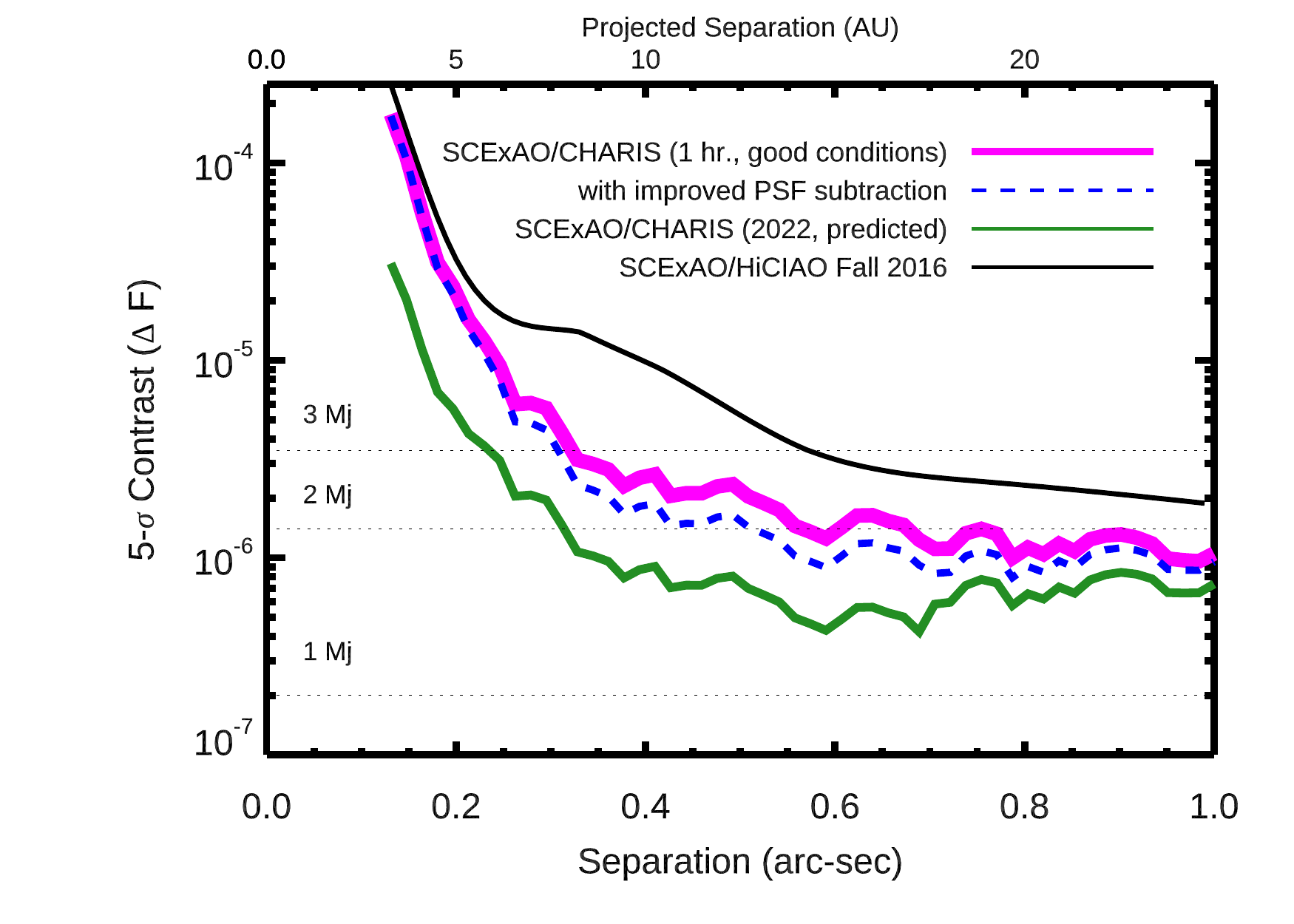}
   \end{center}
   \vspace{-0.3in}
   \caption
   { \label{fig:contrast} 
   SCExAO/CHARIS 5-$\sigma$ contrast curve scaled to one hour integration time (magenta) under good conditions with excellent ($\Delta$PA $>$ 50$^{o}$) parallactic angle motion compared to performance reported in Ref. \citenum{Currie2017} for SCExAO/HiCIAO in Fall 2016, expected performance given recent improvements in PSF subtraction methods, and projected performance in 2022.
   Horizontal dotted lines note the contrasts expected for jovian planets around a 50 Myr-old Sun-like star at a distance of 25 $pc$\cite{Baraffe2003}.}
\end{figure}

\subsection{SCExAO/CHARIS High-Contrast Performance}
Figure \ref{fig:contrast} displays a 5-$\sigma$, 1 hour sequence contrast curve with CHARIS for a 5th magnitude star under good observing conditions and excellent parallactic angle rotation, showing contrasts of 10$^{-5}$, 2$\times$10$^{-6}$, and 10$^{-6}$ 0\farcs{}25, 0\farcs{}4, and 0\farcs{}8, respectively.   Performance degrades for poor field rotation ($\Delta$PA $\lesssim$ 30$^{o}$) or poorer AO corrections.   The data were processed with A-LOCI\cite{Currie2012} first applied to ADI data and then with SDI on the post-ADI residuals.   The SDI PSF subtraction step typically gains a factor of 1.5--2 at mid spatial frequencies; A-LOCI typically outperforms KLIP\cite{Soummer2012} by 20--80\%.   Recent updates to our PSF subtraction algorithm, in particularly more optimal optimization/subtraction zone geometries\cite{Lafreniere2007} in the SDI step suggest an additional gain of 20--40\% at most separations (dashed line).  Simultaneous ADI+SDI PSF subtraction as has been well-demonstrated with SPHERE likewise would provide an additional performance gain\cite{Vigan2015} and is in development.

SCExAO/CHARIS performance is equal to or slightly exceeds that of GPI under good conditions (e.g. 2$\times$10$^{-5}$ at 0\farcs{}25, 5$\times$10$^{-6}$ at 0\farcs{}4\cite{Bailey2016}), while it slightly trails SPHERE's performance under good conditions at small angles (0\farcs{}1--0\farcs{}4)\cite{Vigan2015}.    The near future will see the replacement of AO-188 with a higher order DM, which simulations show should gain a factor of $\sim$ 5 in raw contrast.  This hardware advance coupled with focal-plane wavefront sensing improvements to be commissioned by the end of 2021 should improve performance to $\sim$ 10$^{-6}$ contrast at 0\farcs{}25 and slightly deeper at wider separations (dark green line). 

\begin{figure}[ht]
   \begin{center}
  \centering
   \includegraphics[scale=0.25,clip]{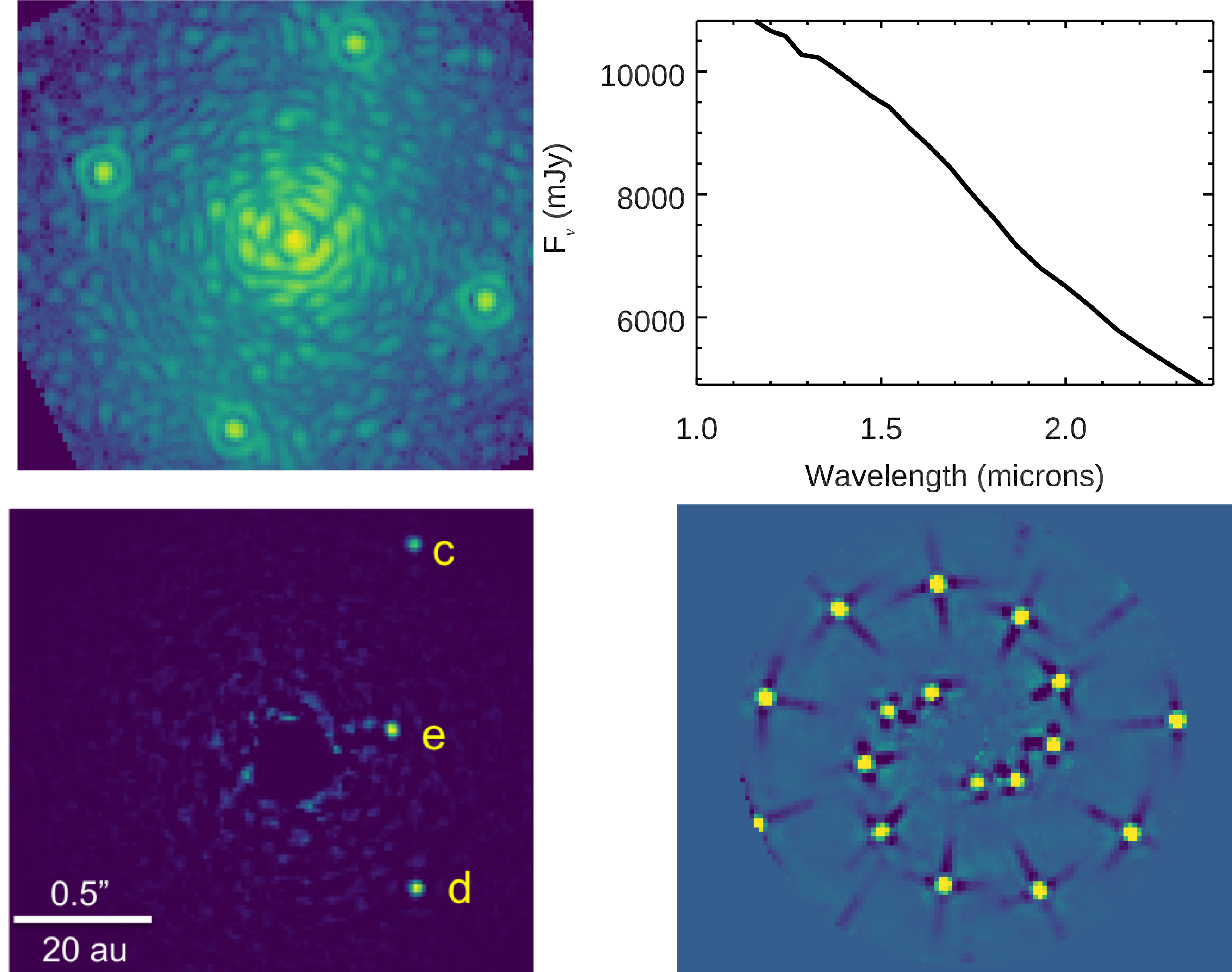}
   \end{center}
   \vspace{-0.1in}
   \caption
   { \label{fig:redsteps} 
   Selected reduction steps in the CHARIS Data Post-Processing Pipeline.  (top-left) A CHARIS data cube after sky-subtraction and precise image registration have been performed.  (top-right) Kurucz model atmosphere for an F8V star binned to CHARIS's spectral resolution and resampled along the CHARIS broadband mode wavelength grid.    (bottom-left) PSF-subtracted image of HR 8799 showing extremely high SNR detections of HR 8799 cde.  (bottom-right) Grid of synthetic L-type planets forward-modeled through our data using the approach of Ref. \citenum{Currie2018a} to simulate signal loss due to A-LOCI in ADI+SDI mode.}
\end{figure}
\section{SCExAO/CHARIS Data Reduction}
\subsection{The CHARIS Data Processing Pipeline: A Walkthrough}
The CHARIS DPP package has been used in nearly all SCExAO/CHARIS science papers thus far
and functions as an end-to-end pipeline, ingesting rectified CHARIS data cubes and producing publication-grade science products: PSF-subtracted images, planet and disk forward-modeling, extracted and calibrated spectra, contrast curves, and basic empirical comparisons.   It is currently written in IDL with plans to translate into Python 3.7 starting in mid/late 2021.   In efforts led by K. Lawson, we are currently beta-testing code to support polarimetric differential imaging with CHARIS's integral field polarimetry mode\cite{vanHolstein2020}.   While the package is currently proprietary, we expect a public version release later in 2021.
Below we describe key reduction steps, a subset of which are shown in Figure \ref{fig:redsteps}.
\begin{itemize}
\item \textbf{Data cube preparation} -- The program \texttt{charis$\_$newobs} creates a directory structure and a parameter file with syntax of \textit{[targetname]+$\_$+[filter(broadband,J,H,K)]+'.info'}, populating it with critical information about the target from SIMBAD and VizieR and default parameters for PSF algorithms.   Next, \texttt{charis$\_$imprep} populates the fits headers with standardized keywords, checks data quality, classifies cubes into science frames and sky frames, and edits the string variables for file number in the \textit{.info} file accordingly.
\item \textbf{Sky subtraction and Image Registration} -- The program \texttt{charis$\_$subtract$\_$sky} performs sky subtraction of each cube element. 
As the sky background level may fluctuate, the user can opt for a simple median sky subtraction or one scaled to the match the background intensity of the reddest $K$-band channels at wide separations.
Each data cube element is registered to coordinate [100,100] (indexed from zero) with \texttt{charis$\_$register$\_$cube} (Figure \ref{fig:redsteps}a).   By default precise registration relies on measuring the centroids of satellite spots in each cube slice, using centroid starting guesses from the middle (usually highest SNR) channel for the bandpass, and refining this estimate using a quadratic functional fit across all high SNR channels.
Registration for unsaturated, unocculated data may use the star itself.  The program also can register a sequence of cubes lacking satellite spots when at least one cube with spots are acquired before/after the sequence by cross-correlating the halo.  
\item \textbf{Spectrophotometric Calibration} -- In \texttt{charis$\_$specphot$\_$cal}, each cube is spectrophotometrically calibrated using the satellite spots and a spectrum of the star (Figure \ref{fig:redsteps}b)\footnote{As a general rule \textbf{\underline{the satellite spots should remain on AT ALL TIMES}}.   Some satellite spots observations are usually required for proper spectrophotometric calibration.   The spots incur negligible loss in speckle suppression.   While SCExAO/CHARIS can take exposures without satellite spots, removing them usually degrades image registration precision and causes significant additional uncertainties in absolute calibration due to transmission and AO performance variations.   Exceptions to this above guideline include programs where science goals require and CHARIS enables unsaturated images of the star.   Likewise, for most programs sky frames before or after science exposures should be taken to remove thermal background at K band.   They also appear to yield small reductions in the background rms at shorter wavelengths.}.   The spectrum may draw from the Kurucz atmosphere library, the Pickles library, or be an empirical spectrum.   By default, flux densities are normalized to one PSF core, although this choice can be overridden.   
\item \textbf{Spatial Filtering} (Optional) \textbf{and PSF subtraction} -- Each data cube slice can be spatially filtered using either a moving-box median or a radial profile (\texttt{charis$\_$imrsub}) before PSF subtraction.   The spatial filtering function provides a quick-look sequence-combined cube and wavelength-collapsed image with the halo suppressed by classical PSF subtraction.    Current available publication-grade PSF subtraction approaches include (for reference star differential imaging, RDI) KLIP, (for ADI) A-LOCI and KLIP, and (for SDI, SDI on the post ADI residuals) A-LOCI: e.g. \texttt{charis$\_$adiklip} (Figure \ref{fig:redsteps}c).   The pipeline nominally adopts default PSF subtraction parameters in the \textit{.info} file, all of which can be overridden at command line. 
\item \textbf{Throughtput Corrections, Forward-Modeling, Spectral Extraction,and Contrast Curves} -- Spectra for a detected source are extracted (\texttt{charis$\_$extract$\_$1d$\_$spectrum}) but must be corrected for signal loss.  The pipeline uses forward-modeling\cite{Currie2018a,Pueyo2016} to determine signal loss due to processing.  In \texttt{charis$\_$aloci(klip)$\_$fwdmod$\_$planet}, forward-modeling is applied at a specific location; similar programs forward-model a grid of synthetic disks.  \texttt{charis$\_$aloci(klip)$\_$attenmap$\_$planet} injects a grid of point sources to yield an attenuation map needed for calculating contrast curves in \texttt{charis$\_$calc$\_$final$\_$contrast}, which produces 1D, 5-$\sigma$ contrast curves in each channel, and in each bandpass.   
\item \textbf{Basic Empirical Analysis} -- The pipeline calculates the spectral covariance at a predefined location\cite{Greco2016} (\texttt{charis$\_$calc$\_$spec$\_$covar}) and compares an extracted spectrum to spectral libraries (\texttt{charis$\_$empbdplanspec}).
\end{itemize}

\textbf{Auto-Reduce} -- To enable quick-look reductions potentially suitable for real-time on-sky analysis, we also incorporate an automatic reduction script (\texttt{charis$\_$autoreduce}) which performs most of these steps with a single command and includes blind source detection and extraction from ADI-reduced data.   

\section{Recent SCExAO/CHARIS Science Results}
 The second (COVID-19 affected) year of full science operations for SCExAO/CHARIS has yielded four peer-reviewed publications from our collaboration \cite{Uyama2020a,Uyama2020b,Lawson2020,Currie2020a}, a complementary publication on SCExAO/VAMPIRES focused on $H_{\rm \alpha}$ imaging searches for protoplanets\cite{Uyama2020c}, and several more CHARIS studies in preparation.   Most results focus on new companions imaged with SCExAO/CHARIS or multi-wavelength characterization of previously known planet-forming disks.   We summarize these published results and several studies in preparation below.
 \subsection{New Companion Detections}
 \begin{figure}[ht]
 \vspace{-0.1in}
   \begin{center}
  \centering
   \includegraphics[scale=0.55,clip]{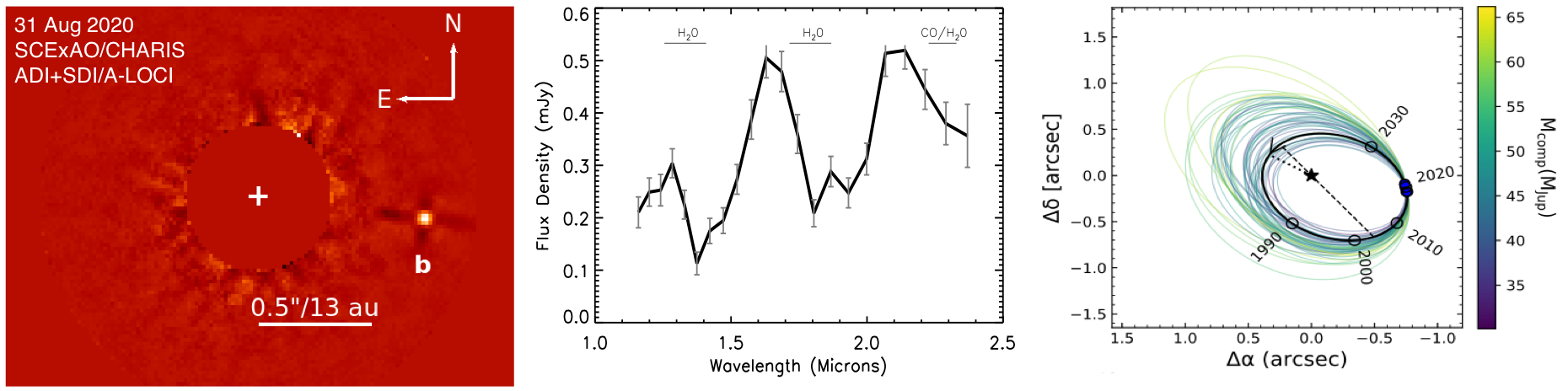}
   \end{center}
   \vspace{-0.1in}
   \caption
   { \label{fig:hd33632} 
   Detection (left), spectrum (middle), and orbital fit/dynamical mass estimate for HD 33632 Ab.   In the wavelength-collapsed CHARIS image, HD 33632 Ab is detected at SNR $\sim$ 68: this particular reduction was not displayed in Ref. \citenum{Currie2020a}.}
\end{figure}

\textbf{HD 33632 Ab}\cite{Currie2020a} -- In SCExAO/CHARIS's first discovery paper, we presented the detection of HD 33632 Ab, a substellar companion to a nearby mature Sun-like star, HD 33632 Aa imaged at a projected separation of $\sim$ 20 au (Figure \ref{fig:hd33632}).   The companion's colors and near-IR spectrum are best-matched by field objects at the L/T dwarf transition, where clouds/dust in substellar atmospheres dissipate/sink below the photosphere.   The companion may be a particularly useful reference point for understanding the first directly imaged exoplanets, as its colors overlap with and its temperature is just slightly exceeds those of the young exoplanets HR 8799 cde.  

Unlike nearly all other ``benchmark'' L/T transition substellar objects, HD 33632 Ab has both a high-quality near-IR spectrum AND a direct dynamical mass measurement because of the astrometric acceleration it induces on its host star, identified from the Hipparcos and Gaia satellite data.   Assuming it is solely responsible for the primary's acceleration, HD 33632 Ab's inferred mass is $\sim$ 46 $M_{\rm J}$ $\pm$ 8 $M_{\rm J}$, which is comfortably above the deuterium-burning limit nominally separating planets from brown dwarfs.   However, its mass, mass ratio, and separation are comparable to multiple companions near the nominal planet to brown dwarf boundary.   Its eccentricity must be low and may be more similar values for bona fide directly imaged planets than companions identified as brown dwarfs based on their masses.   

Future follow-up CHARIS data at higher resolution will refine its spectral properties and more precisely determine its mass, assuming that it is the only massive object accelerating its host star.   Deep follow-up CHARIS data could identify any hitherto unseen companions at small separations; however, due to the system's age, JWST/NIRCam thermal IR imaging likely will provide a more sensitive search for cool companions lower in mass than HD 33632 Ab.   Multi-band thermal IR imaging could further probe filters carbon chemistry and absorption from other species (e.g. CO$_{\rm 2}$).   While the ground could provide some of these data (e.g. at 3.1 and 3.3 $\mu m$), high thermal backgrounds likely make a high SNR detection and precise photometry at $M_{\rm s}$ (i.e. $\sigma$ $\lesssim$ 0.1 mag) implausible with current facilities (see \cite{Galicher2011}) .   Here again, JWST is natural complement, as NIRCam imaging -- e.g. 3.6 $\mu m$ and 4.3--4.8 $\mu m$ -- will be essential for better characterizing this system and providing a context for HR 8799 bcde and other L/T transition planet-mass objects.

\begin{figure}[ht]
   \begin{center}
  \centering
   \includegraphics[scale=0.525,trim=0mm 10mm 0mm 10mm,clip]{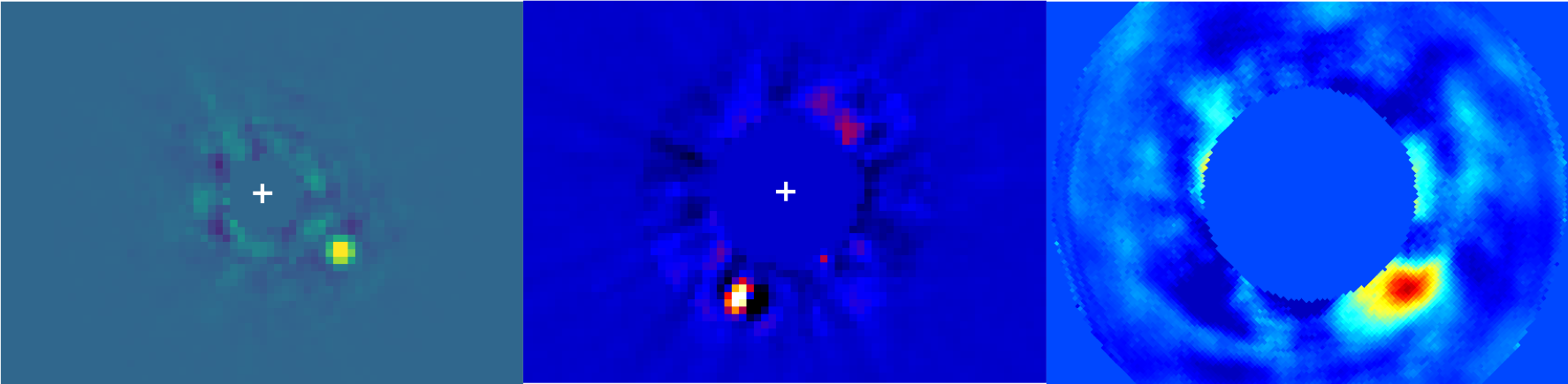}
   \end{center}
   \vspace{-0.1in}
   \caption
   { \label{fig:newdisc} 
   Three additional new discoveries with SCExAO/CHARIS. 
   }
\end{figure}

\textbf{Other Systems} -- In addition to HD 33632 Ab, SCExAO/CHARIS has yielded over a half-dozen newly-detected (likely) companions, three of which are shown in Figure \ref{fig:newdisc}.   The most unique of these shown in a new candidate infant planet detected in over half a dozen CHARIS data sets (right panel) (Currie et al. in prep).   While distinguishing between bona fide protoplanets and disk features is notoriously challenging, the signal's spectrum, astrometry, and other features provide good evidence for a protoplanet interpretation.   In addition to these objects and several other more massive stellar companions, SCExAO/CHARIS also has a multiple data set detection of a candidate fully-formed directly imaged planet with properties plausibly similar to HR 8799 bcde (T. Currie in prog.).   This detection is not shown in this paper.  

 \subsection{Characterization of Planet-Forming Disks}
\begin{figure}[ht]
   \begin{center}
  \centering
   \includegraphics[scale=0.45,clip]{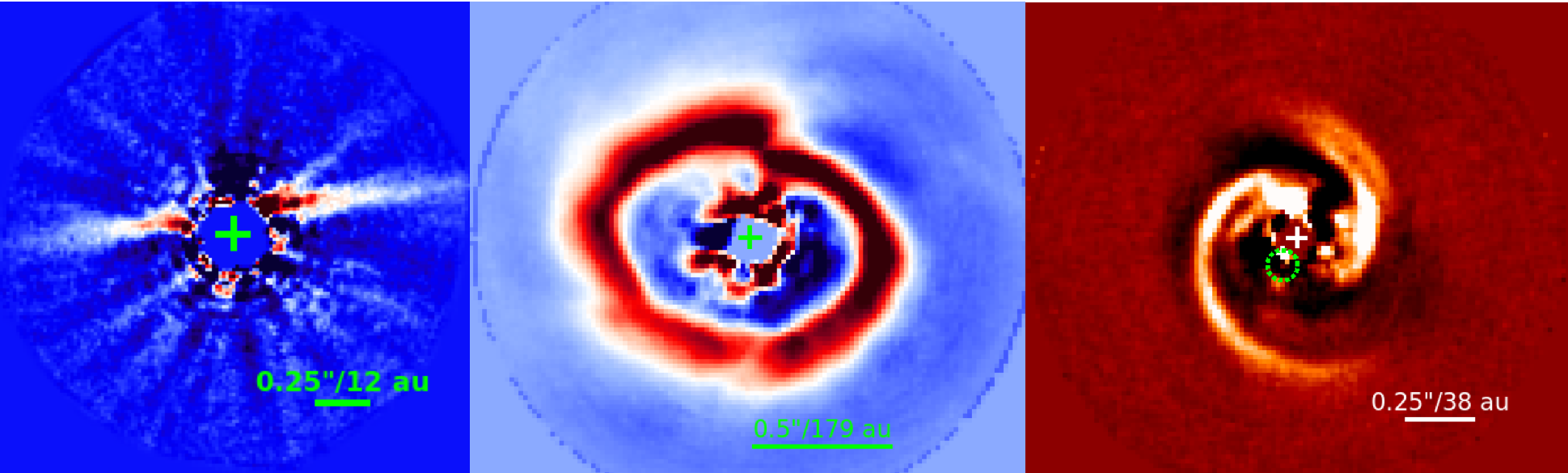}
   \end{center}
   \vspace{-0.1in}
   \caption
   { \label{fig:diskgallery} 
   Planet-forming disks resolved with SCExAO/CHARIS.   (left) The HD 15115 (``Blue Needle") debris disk\cite{Lawson2020}.  (middle) The broken ring resolved around HD 34700 A\cite{Uyama2020a}.   (right) Unpublished image of the MWC 758 protoplanetary disk.   The previously claimed protoplanet candidate (dashed circle) is not detected.
   }
       \vspace{-0.2in}
\end{figure}

 \textbf{HD 15115}\cite{Lawson2020} -- The first discovery with SCExAO -- then coupled with the HiCIAO infrared camera -- was a resolved $H$-band image of the bright debris disk around the 5--10 $Myr$ old star, HD 36546 \cite{Currie2017}.   SCExAO/CHARIS imaging of another debris disk -- the cold debris disk around HD 15115, also known as the ``Blue Needle"\cite{Kalas2007} -- provided a multi-wavelength look at Kuiper belt-like structures to constrain dust composition and scattering properties(Figure \ref{fig:diskgallery}a).   We detected the HD 15115 disk down to separations of $\sim$ 0\farcs{}2, a factor of 3--5 smaller than previous studies.   
 
We recover an east-west disk brightness asymmetry previously seen at wider separations and at other wavelengths\citenum{Kalas2007}.   However, the intrinsic disk colors appear red at small separations, in contrast to the disk's blue colors at wide separations.  While a SPHERE study suggested a misaligned two-ring geometry\cite{Engler2019}, our more sensitive data showed that a single ring with a Hong-like scattering phase function fit the data well.   

\textbf{HD 34700 A}\cite{Uyama2020a} -- We resolve the bright broken ring around HD 34700 A and recover multiple spirals arms previously seen in polarimetric data\cite{Monnier2019}(middle panel).  Geometric albedos derived from the ring's surface brightness profile point to a large scale height or copious submicron-sized dust at position angles between $\sim$ 45$^{o}$ and 90$^{o}$.   A stellar flyby or envelope infall may explain the spirals' very large pitch angles.

\textbf{MWC 758} -- In an unpublished work, we resolved multiple spiral arms in the MWC 758 protoplanetary disk (right panel).   The presence of spiral density waves has been attributed to massive, hidden planets in this system\cite{Muto2012}.   Two studies have identified candidate protoplanets that may be connected to these spirals: a bright inner companion at $\rho$ $\sim$ 0\farcs{}11\cite{Reggiani2018} and a much fainter outer one at $\rho$ $\sim$ 0\farcs{}62\cite{Wagner2019}, both detected at $L_{\rm p}$.   CHARIS data well resolve MWC 758's spiral arms, both of which appear to have two components, and provide constraints on other disk material and bright protoplanets down to 0\farcs{}05.    Our data might not be sensitive enough to assess the nature of the faint, wide-separation candidate protoplanet.

Analysis of these data provide evidence against the protoplanet reported in Ref. \citenum{Reggiani2018}: no astrophysical is seen at its predicted location (dashed circle).  To further assess its nature, we downloaded and examined the Ref. \citenum{Reggiani2018} images provided by \textit{VizieR}.   While Ref. \citenum{Reggiani2018} report an SNR of $\sim$ 5 and 6 for the two epochs, using our adopted computation for SNR, we instead find that the detections are not statistically significant compared to other signal at the same angular separations: SNRs $\sim$ 1.8 and 2.0, respectively

This result reaffirms the significant challenge of distinguishing between bona fide protoplanets and disk features\cite{Currie2019a}.  It also a case study of complications with interpreting the nature of signals near the diffraction limit that emerge after PSF subtraction techniques are used to (imperfectly) whiten highly non-Gaussian noise.

\section{Future Directions}
\subsection{Technical Advances}
A detailed overview of recent and upcoming technical improvements to SCExAO is discussed in various complementary SPIE submissions.   A subset of these upgrades are discussed below, particularly focusing on wavefront sensing/control advances.   
\begin{itemize}
    \item \textbf{Replacement and Upgrade of AO-188} - We plan to replace Subaru's venerable facility AO system, which uses a curvature wavefront sensor driving a DM with only 188 actuators.   AO-188's low actuator density across the telescope pupil allows a modest correction of atmospheric turbulence and exhibits strong vibration modes precluding full-speed operation\cite{Oya2006}.   Its successor will operate with a much faster (2 kHz) and far higher-order DM (3200 actuators).   Our plan is to upgrade its wavefront sensor from the current 188-element avalanche photodiode array to an EMCCD.   The camera upgrade will allow the ``AO-3000" to substantially reduce both the fitting error and temporal bandwidth error, yielding $H$ band Strehl ratios approaching 85\% and a reduction in the PSF halo by nearly a factor of 5.    
    \item \textbf{Self-adjusting modal AO control} -  Work is in progress to implement an automatic modal gain algorithm\cite{Deo2019}, which continuously tracks the adequate gain for minimum variance control of all AO modes, while compensating for the varying sensititivity of the PyWFS (a nonlinear effect depending on seeing, source brightness, etc.). Daytime tests have shown adequate control with varying turbulence conditions for up to 400 DM modes. On-sky testing is ongoing to improve the robustness of the algorithm.
    
    \item \textbf{AO optimization through reinforcement learning} - We are developing an algorithm - Dr WHO (Direct Reinforcement Wavefront Heuristic Optimization) aiming to correct on-sky the static, quasi-static and dynamic non-common path aberrations by updating the reference of the pyramid wavefront sensor. The new references are chosen based on the lucky imaging of the focal plane camera. See SPIE paper 11448-255 for more details. 
    
    
    \item \textbf{Focal-Plane Wavefront Sensing with MEC} - The MKID Exoplanet Camera (MEC) is a noiseless, ultra-cooled photon-counting detector able to measure the energy and wavelength of every photon. and can function both as a science instrument and a focal-plane wavefront sensing sensor integrated with SCExAO\cite{Walter2020}.  MEC operates from $z$ to $J$ band (0.8--1.4 $\mu m$) and often runs in parallel with CHARIS observations, taking $Y$ band light.    Operating at the focal plane, MEC is now demonstrating the ability to perform slow speckle nulling capable of cancelling bright, ``slow" quasi-static speckles that evolve on tens of minutes timescales.   Given its exceptional sensitivity and fast readout ($>$ 2 kHz), MEC can also cancel fast ($\tau_{\rm o}$ $\sim$ 10 ms) atmospheric speckles and generate an extremely deep-contrast dark hole (DH).    In both cases, MEC will enable significantly deeper raw contrasts beyond those provided by AO-3000 and SCExAO alone.
    
    \item \textbf{Linear Dark Field Control} -- After generating a dark hole using a method like speckle nulling with MEC or another focal-plane camera, \textit{Linear Dark Field Control}(LDFC)\cite{Miller2017} could ``freeze" the residual DH state without the need for probing.   
    Laboratory tests of LDFC at raw contrasts ($\sim$ 5$\times$10$^{-7}$) relevant for imaging reflected-light planets from ground-based telescopes demonstrated its ability to correct for a range of phase perturbations with improved efficiency relative to classical speckle nulling\cite{Currie2020b}.   Tests of LDFC at slightly milder contrasts using SCExAO's internal source and simulated turbulence as well as preliminary on-sky tests were extremely promising (K. Miller et al. 2020 submitted; S. Bos 2021 in prep.).   
    
    \item \textbf{Alternating Speckle Grid for High-Precision Spectroscopy} -- 
    To achieve higher-precision spectroscopy without the need for very bright satellite spots, we have developed an alternating speckle grid strategy (Figure \ref{fig:altgrid}).
The underlying incoherent background is subtracted by taking two exposures with alternating speckle patterns \cite{sahoo2020precision}. 
Current on-sky photometric and astrometric precision (obtained by measuring the flux ratio and relative position between calibration spots) for a 30 minute observation time with is 0.3$\%$ and 1.7mas respectively.  We plan to begin support of this mode within the CHARIS DPP in 2021.

\begin{figure}
\centering
\begin{subfigure}{0.35\textwidth}
  \centering
  \includegraphics[width=\linewidth]{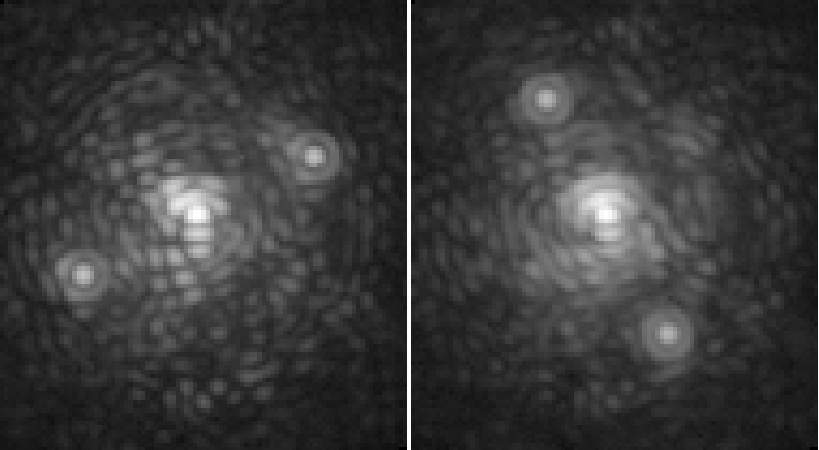}
  \caption{}
  \label{fig:sub1}
\end{subfigure}%
\begin{subfigure}{0.35\textwidth}
  \centering
  \includegraphics[width=\linewidth]{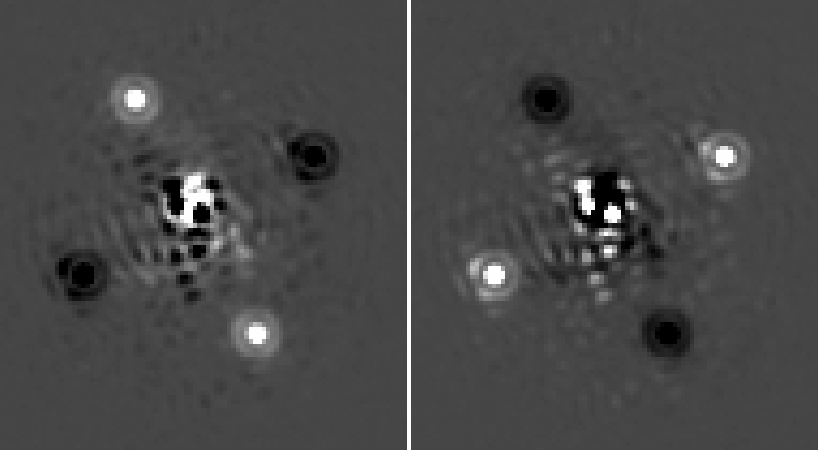}
  \caption{}
  \label{fig:sub2}
\end{subfigure}
\vspace{-0.05in}
\caption{On-sky images of two consecutive reduced CHARIS data slices of $\theta$ Hydrae with satellite spots (a) Incoherent speckles (without background subtraction) and (b) Alternating technique; after background subtraction obtained from CHARIS at 1.63 $\mu m$ with the two speckle patterns at $\rho$ $\sim$ 11$\lambda/D$.}
\label{fig:altgrid}
\vspace{-0.2in}
\end{figure}

    
\end{itemize}

\subsection{SCExAO/CHARIS Direct Imaging Survey}
   

We have recently initiated an exoplanet direct imaging survey focused on targeting stars showing indirect evidence for a planet from astrometry.  The low yield of \textit{blind} direct imaging surveys show that exoplanets directly detectable with current instruments are rare at 5--100 au\cite{Nielsen2019}.  Companions detected by these surveys are typically more than 2--5 $M_{\rm J}$ and orbit beyond $\sim$ 3 au, where the jovian planet frequency peaks.   Limited sample sizes and sparse coverage of ages, temperatures, and surface gravities impedes our understanding of the atmospheric evolution of gas giant planets.  However, targeted searches focused on stars showing evidence of gravitational pulls from massive planets could improve survey yields.

The {\it Hipparcos}-{\it Gaia} Catalog of Acclerations (HGCA) provides absolute astrometry for 115,000 nearby stars, including those with clear dynamical evidence for unseen massive companions\cite{Brandt2018}.   Accelerations derived from the HGCA can provide dynamical masses of known imaged exoplanets and low-mass brown dwarfs independent of luminosity evolution models and irrespective of uncertainties in stellar ages\cite{Brandt2019}.    

So far, \textbf{{our survey has been a resounding success}}.   This approach has already yielded the discovery of HD 33632 Ab and and numerous other low-mass companions and other planet/brown dwarf candidates.   Thus far, our detection rate of (candidate) companions is $\sim$ 30\%.
For target selection, we have explored identifying accelerating stars 1) in young moving groups and 2) among the field.   Thus far, we favor selecting field objects, as they are typically nearer and have higher quality astrometry than stars in many moving groups or associations: moving group members with statistically signficant accelerations in HGCA are often stars with K or M dwarf companions (see also Ref. \citenum{DeRosa2019}).   Other objects -- e.g. white dwarfs -- may also be responsible for accelerations identified from astrometric data\cite{Bonavita2020}.   Age estimates for field stars are more uncertain than for moving group members.   However, this has negligible impact on our derived masses since Hipparcos and Gaia absolute astrometry and SCExAO/CHARIS relative astrometry provide dynamical masses.

The Gaia Early Data Release 3 and full Data Release 3 (scheduled for 2022) will provide additional precise astrometry for nearby stars and will reveal more systems showing evidence for hitherto unidentified but imageable planets.   Combining these new measurements with previous HGCA absolute astrometry for SCExAO/CHARIS-discovered planets and brown dwarfs will yield an even more robust mapping between a substellar object's atmospheric properties and its bulk properties (mass).

\acknowledgments 
\indent The authors acknowledge the very significant cultural role and reverence that the summit of Mauna Kea holds within the Hawaiian community.  We are most fortunate to have the opportunity to conduct observations from this mountain.   We support and endeavor to contribute to respectful, effective stewardship of cultural, natural, and scientific resources that properly honors these lands.   

\indent We acknowledge the critical importance of the current and recent Subaru Telescope daycrew, technicians, support astronomers, telescope operators, computer support, and office staff employees, especially during the challenging times presented by the COVID-19 pandemic.  Their expertise, ingenuity, and dedication is indispensable to the continued successful operation of these observatories.  
\\
\indent We thank the Subaru Time Allocation Committee for their generous support of this program.  TC was supported by a NASA Senior Postdoctoral Fellowship and NASA/Keck grant LK-2663-948181.   TB gratefully
acknowledges support from the Heising-Simons foundation and from NASA under grant \#80NSSC18K0439. 
\\
\indent The development of SCExAO was supported by JSPS (Grant-in-Aid for Research \#23340051, \#26220704 \& \#23103002), Astrobiology Center of NINS, Japan, the Mt Cuba Foundation, and the director's contingency fund at Subaru Telescope.  CHARIS was developed under the support by the Grant-in-Aid for Scientific Research on Innovative Areas \#2302. 
\bibliography{report} 
\bibliographystyle{spiebib} 

\end{document}